\documentclass[journal]{vgtc}                     

\onlineid{8382}

\vgtccategory{Research}

\title{ProGram: A Grammar for Progressive Computation Pipelines}

\author{%
  Seunghoon Jung and
  Jaemin Jo
}

\authorfooter{
  \item Seunghoon Jung is with Sungkyunkwan University.
  E-mail: shoon125@g.skku.edu
  \item Jaemin Jo is with Sungkyunkwan University and is the corresponding author.
  E-mail: jmjo@skku.edu
  	
}

\abstract{%
Progressive Visual Analytics (PVA) systems have demonstrated the effectiveness of time-bounded, incremental computation for exploring large-scale data. However, our survey reveals that existing PVA implementations are often difficult to reuse and extend, as their progressive behaviors are typically realized through ad hoc, hard-coded mechanisms distributed across servers and clients.
To address this limitation, we present \ours, a grammar for progressive computation pipelines. Informed by a comprehensive analysis of prior concepts and design practices in existing PVA systems, \ours is designed to support core progressive features while reducing implementation burden.
Furthermore, we extend the design space by introducing novel mechanisms, such as lockstepping, forking, and temporary prioritization, to better support diverse usage scenarios. 
Finally, we demonstrate the expressiveness and coverage of \ours by reproducing and extending existing progressive visualization systems, presenting a gallery of progressive computation pipelines, and discussing the grammar coverage.
}


\teaser{
  \centering
  \includegraphics[width=0.95\linewidth, alt={}]{figs/teaser.png}
  \caption{
  \ours in use.
  (A) The reproduced ProReveal interface.
(B) The DimXplorer interface, reproduced and extended with dynamically created views, where the clustering and PCA views progress in lockstep.
(C) Two gallery examples: iteration-based chunking with $t$-SNE (C1) and lockstepping between two views (C2).
 }
  \label{fig:teaser}
}

\graphicspath{{figs/}{figures/}{pictures/}{images/}{./}} 

\usepackage{tabu}                      
\usepackage{booktabs}                  
\usepackage{lipsum}                    
\usepackage{mwe}                       
\usepackage{ccicons}                   

\usepackage{mathptmx}                  

\usepackage{mathptmx}                  
\usepackage{kotex}
\usepackage{mathtools}

\usepackage{newfloat}
\usepackage[newfloat]{minted}
\usepackage{caption}
\usepackage{tabularx}
\usepackage{makecell}

\usepackage{xcolor}
\usepackage{tikz}
\usepackage[most]{tcolorbox}

\definecolor{schemaBlue}{HTML}{DCE8F1}
\definecolor{schemaGreen}{HTML}{E1EAD8}
\definecolor{schemaYellow}{HTML}{F2E9CF}
\definecolor{schemaPurple}{HTML}{E7DDEA}
\definecolor{schemaRed}{HTML}{F1DDDA}

\definecolor{pipelineColor}{HTML}{F6E3B4}
\definecolor{binningColor}{HTML}{DCE8F1}
\definecolor{reducingColor}{HTML}{DCEBD8}

\definecolor{fieldColor}{HTML}{A33A3A}
\definecolor{typeColor}{HTML}{3F5F7A}
\definecolor{commentColor}{HTML}{4F7A55}

\definecolor{queryColor}{HTML}{DCE8F1}
\definecolor{progressionColor}{HTML}{E5E1F0}
\definecolor{wranglingColor}{HTML}{F6E3B4}
\definecolor{binningColor}{HTML}{E7DDEA}
\definecolor{reducingColor}{HTML}{DCEBD8}

\definecolor{fieldColor}{HTML}{A33A3A}
\definecolor{typeColor}{HTML}{3F5F7A}
\definecolor{literalColor}{HTML}{333333}

\newcommand{\qfield}[1]{\textcolor{fieldColor}{#1}}
\newcommand{\qtype}[1]{\textcolor{typeColor}{#1}}
\newcommand{\qlit}[1]{\textcolor{literalColor}{#1}}

\definecolor{transformingColor}{HTML}{F3DFB8} 
\definecolor{filteringColor}{HTML}{E8D8C8}    

\newcommand{\synkey}[1]{\textcolor{blue!65!black}{#1}}
\newcommand{\synstr}[1]{\textcolor{orange!75!black}{#1}}
\newcommand{\synnum}[1]{\textcolor{purple!70!black}{#1}}
\newcommand{\synkw}[1]{\textcolor{violet!70!black}{#1}}
\newcommand{\synfun}[1]{\textcolor{teal!65!black}{#1}}

\newcommand{\schematag}[2]{%
  \tikz[baseline=(tag.base)]{
    \node[
      anchor=base,
      fill=#1,
      draw=black!12,
      line width=0.3pt,
      rounded corners=2pt,
      inner xsep=2.5pt,
      inner ysep=0.7pt,
      font=\ttfamily\bfseries\scriptsize
    ] (tag) {\strut #2};
  }%
}

\newcommand{\code}[1]{{\footnotesize\ttfamily #1}}

\newcommand{\ours}{\textit{ProGram}\xspace}

\usepackage{xspace}

\setminted{fontsize=\fontsize{6.5}{7.5}\selectfont,
  numbersep=6pt,
  xleftmargin=1em,
  breaklines,
  frame=single
  }

\begin{document}

\crefname{listing}{Code}{Codes}
\Crefname{listing}{Code}{Codes}


\firstsection{Introduction}

\maketitle

Progressive Visual Analytics (PVA) mitigates the latency of analyzing large datasets by decomposing a long-running computation into a sequence of partial results.
Rather than visualizing the entire dataset at once, PVA visualizes intermediate results and progressively refines them over time.
This keeps users engaged during long-running analyses, allows earlier decisions from partial results, and can reduce computational cost by terminating unnecessary processing.
Motivated by these benefits, a wide range of PVA systems and user interfaces have been developed~\cite{Zhang-PEViz, Ulmer-ProBGP, Jo-ProReveal, SampleActoin-trust-me-im-partially-right-2012-Fisher, Prosample-Tailorable-Sampling-for-PVA-2024-Hografer, DimXplorer-Desinging_progressive-and-interacitve-process-for-high-dimensional-data-analysis-2017-turkay, IncVisage-Ive-seen-enough-2017-Rahman-Ronitt, MolSieve-PVA-for-molecular-dynamics-simulation-2024-Hnaatyshyn}.
Each of them, however, is dedicated to a specific dataset, task, or analytical workflow, with its progressive behavior tightly coupled to the underlying data processing logic, and is therefore built from scratch on a bespoke codebase.

Although these systems are built independently, they share a common progressive computation pipeline, in which queries are issued to the server, computations are executed progressively, partial results are delivered to the client, and progression and scheduling are continuously managed in response to user interaction.
Prior surveys and books~\cite{A-Survey-on-Progressive-Visualization, Angelini-A-review-and-characterization-of-PVA, The-human-user-in-PVA-2019-Micallef, PDABook} have characterized this pipeline by contrasting PVA with earlier computational paradigms, identifying interaction mechanisms, and examining user roles and tasks.
These characterizations, however, have not been translated into reusable building blocks that developers can assemble a system from.
Each system therefore implements the shared pipeline on its own, splitting its logic across the server and the client and coordinating the two through ad hoc mechanisms tailored to a specific analytical setting.

The InfoVis community has addressed comparable situations with formal grammars, which conceptualize a design space and expose it as declarative specifications backed by a shared runtime, for general visualization design~\cite{vega-2016-Satyanarayan-Jeffery, vega-lite}, comparative view composition~\cite{lyi2020-comparative-layouts-revisited-design}, density map rendering~\cite{Declarative-rendering-density-map-2019-Jo}, and statistical model checks~\cite{vmc-statistical-model-check-2025-Guo}.
Motivated by their success, we present \ours, a grammar for the progressive computation pipeline, grounded in a survey of 21 articles analyzed from a developer perspective, together with a runtime that executes its specifications across the server and the client.
\ours specifies how progressive computations are executed, moving coordination logic from hard-coded server-side implementations into client-side declarative specifications, so that the server runs only a generic engine and developers can build a progressive analysis without writing application-specific server code.
We demonstrate the expressiveness of \ours by reproducing and extending two existing PVA systems, ProReveal~\cite{Jo-ProReveal} and DimXplorer~\cite{DimXplorer-Desinging_progressive-and-interacitve-process-for-high-dimensional-data-analysis-2017-turkay}, and by presenting a gallery of eleven pipelines, which we release publicly\footnote{The code will be released upon acceptance}.

In summary, our contributions are three-fold:
\begin{enumerate}
    \item We survey prior PVA systems, toolkits, and grammars from an implementation-oriented perspective, revealing their limitations in reusability, extensibility, and composability.
    \item We present \ours, a grammar for the progressive computation pipeline, with novel execution control mechanisms such as lockstepping, forking, and temporary prioritization.
    \item We provide a runtime that executes \ours specifications across the server and the client, and demonstrate its utility by reproducing, extending, and creating progressive computation pipelines.
\end{enumerate}
\section{Related Work}

\subsection{Progressive Visual Analytics}
Progressive Visual Analytics (PVA) has emerged as a paradigm for exploring large-scale datasets.
Stolper et al.~\cite{PVA-Progressive-Insight-2014-Stolper} introduced PVA to sustain user attention during long-running computations, where sequential processing cannot deliver interactive response times.
Unlike traditional approaches that visualize results only after the entire dataset has been processed, PVA chunks either the data or the computation to obtain intermediate results and visualizes them incrementally~\cite{A-Survey-on-Progressive-Visualization}.
Rather than producing a single approximate result, PVA continuously refines the same visualization and lets users terminate the process once the remaining uncertainty becomes acceptable~\cite{Angelini-A-review-and-characterization-of-PVA, ProInterest-Combining-Degree-of-Interest-and-PVA-2023-Hografer, The-human-user-in-PVA-2019-Micallef}.
Users can thus gain insights or make decisions early, without waiting for the full computation to complete.

Building a PVA system poses difficulties at two levels.
At the algorithmic level, transforming a conventional algorithm into an incremental or progressive form is often non-trivial.
Welford's online variance algorithm, for example, reformulates batch variance computation as a numerically stable recurrence that updates the mean and variance without retaining all previous observations~\cite{welford-algorithm-1962-welford}, and progressive $t$-SNE incrementally updates an embedding as new data are processed while preserving the stability of previously positioned points~\cite{panene-2018-jo}.
At the system level, even when such algorithms are available, turning them into an analysis that others can adopt remains challenging.
We refer to this distance between a progressive algorithm and a reusable progressive analysis as the \textit{implementation gap} of PVA.

We aim to address the system level, for two reasons.
First, the algorithmic side is already served by a growing body of work, whereas how such an analysis is authored and shared has received far less attention.
Second, the two levels are complementary rather than competing: an abstraction that closes the implementation gap stays agnostic to which progressive algorithm it drives, and therefore inherits every advance made at the algorithmic level.
What is missing is a shared vocabulary in which a progressive analysis can be written down once and reused, rather than a system that must be rebuilt for each analysis.
We further scope this vocabulary to the computation behind a progressive visualization, as the visual side is already served by a large body of work on visualization grammars.

\subsection{Surveys on Progressive Visual Analytics}

To understand how PVA systems have been built and to locate the implementation gap within this body of work, we began with three prior surveys~\cite{A-Survey-on-Progressive-Visualization, Angelini-A-review-and-characterization-of-PVA, The-human-user-in-PVA-2019-Micallef}, which examine PVA from conceptual, human-centered, and visualization-centered perspectives, respectively.
Angelini et al.~\cite{Angelini-A-review-and-characterization-of-PVA} characterize the concept of PVA by contrasting it with instantaneous and monolithic visual analytics, and derive a set of practical requirements and guidelines for PVA systems.
Micallef et al.~\cite{The-human-user-in-PVA-2019-Micallef} shift the focus to the human user, providing a human-centered characterization that identifies user roles (i.e., Observer, Searcher, and Explorer), the analytic tasks and foci associated with each role, and four classes of cognitive bias that may arise when interpreting progressive results.
The most recent survey, by Ulmer et al.~\cite{A-Survey-on-Progressive-Visualization}, concentrates on the design space of progressive visualization and is accordingly scoped to work that incorporates at least one form of it.
It organizes this space along several dimensions, including visualization type, mode of progressive processing (e.g., data chunking or process chunking), data domain, and visual update pattern.

Together, these surveys characterize what PVA is, who uses it, and how progressive results are displayed, but none of them examines how progressive systems are implemented.
We therefore shift the focus from the user, who operates a PVA system, to the developer, who implements such a system or converts an existing monolithic one into a progressive one.
To assess the extent to which current PVA systems can be adopted and reused in practice, we survey PVA systems, toolkits, and grammars from an implementation-oriented perspective, and discuss the challenges developers face when adopting and extending them.

\subsection{Visualization Grammars}
Authoring visualizations often requires substantial effort to manage low-level implementation details.
Grammar-based approaches address this by abstracting visualization design into declarative specifications.
Vega-Lite~\cite{vega-lite}, for example, lets authors specify visual encodings without attending to rendering details, reducing the effort of building and revising visualizations.
VMC~\cite{vmc-statistical-model-check-2025-Guo} applies the same idea to a narrower domain, expressing visualizations for statistical model checking as a grammar.
We pursue the same goal for the progressive computation pipeline.

Our work builds upon prior efforts toward reusable development tools for PVA, of which P5~\cite{kwan-liu-P5}, ProgressiVis~\cite{Fekete-ProgressiVis}, and ProVega~\cite{provega-grammar-ease-prototyping-2026-filosa-marco} are closest to ours.
P5 is a web-based toolkit that provides a declarative grammar for bar charts, histograms, binned scatterplots, and geographic maps, and additionally supports declarative specification of user interactions such as data cube-based filtering.
ProgressiVis is a Python framework of progressive computation operations, typically used within Jupyter notebooks.
It adopts a scheduler-based execution model in which modules run for a short time quantum and then yield control, allowing analysts to monitor intermediate outputs and steer computations interactively.
ProVega extends Vega-Lite with declarative constructs for progression, monitoring, and user control, while allowing external progressive backends to be connected through hooks.

Our survey, however, reveals a fundamental discrepancy between these tools and the PVA systems deployed in practice.
Most real-world systems adopt a distributed execution model that separates computation on the server from visualization on the client, yet the existing tools provide limited support for this architecture. 
ProgressiVis is tightly coupled to the Jupyter notebook environment, making it ill-suited to the web-based interfaces that predominate in practice, whereas P5 assumes that all data processing occurs locally in the browser, which is incompatible with scalable server-side processing.
ProVega can connect progressive visualizations to external computation backends, but does not itself provide a server-side computation runtime or abstractions for coordinating computations across the client-server boundary.
\ours aims to remove this constraint: an analysis is written once, independently of where its computation runs, and remains a specification that others can run and adapt.

\textbf{Positioning of our work.}
Existing approaches primarily abstract either progressive computation on the server side or the visualization authoring on the client side.
\ours instead addresses the layer between them: how a client issues a query to the server and receives its partial results, how multiple such queries are coordinated together, and how the user steers the ongoing computation.
\ours is therefore not tied to a specific visualization grammar or chart type; it provides the basis on which visualizations and interactions are served, which existing approaches leave to ad hoc, application-specific implementations.
\section{Implementation Practices in PVA Systems}

\begin{table*}[t]
\centering
\scriptsize
\setlength{\tabcolsep}{3pt}
\renewcommand{\arraystretch}{1.15}
\resizebox{\textwidth}{!}{%
\begin{tabular}{l *{13}{c}}
\toprule
&\multicolumn{1}{c}{}
& \multicolumn{1}{c}{Architecture}
& \multicolumn{6}{c}{Analytical Coverage}
& \multicolumn{5}{c}{Progressive Control by the User} \\
\cmidrule(lr){3-3}\cmidrule(lr){4-9}\cmidrule(lr){10-14}
Systems
& \makecell[c]{Open-source}
& \makecell[c]{Server / Client\\Language}
& \makecell[c]{Selection /\\Filtering}
& \makecell[c]{Count /\\Sum}
& \makecell[c]{Min /\\Max}
& \makecell[c]{Clustering}
& \makecell[c]{Projection}
& \makecell[c]{Others}
& \makecell[c]{Concurrent\\Queries}
& \makecell[c]{Pausing /\\Resuming}
& \makecell[c]{Scheduling}
& \makecell[c]{Query \\Prioritization}
& \makecell[c]{Ad-hoc \\Querying} \\
\midrule
SampleAction\cite{SampleActoin-trust-me-im-partially-right-2012-Fisher} %
&X & ? / ? 
& O & O & X & X & X & -
& $N$ & O & ? & X & O \\

Vizdom\cite{vizdom-2015-crotty}
&X & ? / ? 
& O & O & X & X & X & Regression
& $N$ & X & ? & X & O \\

InsightsFeed\cite{Steering-the-Craft-Insightfeed-2017-Badam-Jean-Daniel} %
&X & Python / JS 
& O & O & X & $k$-means & $t$-SNE & -
& 4 & O & ? & X & X \\

IncVisage\cite{IncVisage-Ive-seen-enough-2017-Rahman-Ronitt}
&O & Python + C++ / JS 
& O & O & X & X & X & Sampling
& 1 & O & - & - & X \\

SwiftTuna\cite{Swifttuna-2017-Jo}
& X & \makecell[c]{Python / JS}
& O & O & O & X & X & -
& $N$ & O & RR / FIFO & O & O \\

DeepEyes\cite{Deepeyes-2018-Pezzotti}
& X & \makecell[c]{- / JS + C++}
& O & X & X & X & $t$-SNE & DL model metrics 
& 1 & X & - & - & X \\

PPMT\cite{PPMT-Progressive-sequential-pattern-mining-Raveneau}
& O & Java / JS
& O & X & X & X & X & Pattern mining
& 1 & X & - & - & X \\ 

DimXplorer\cite{DimXplorer-Desinging_progressive-and-interacitve-process-for-high-dimensional-data-analysis-2017-turkay}
& X & - / ?
& O & X & X & $k$-means & PCA & -
& 2 & O & ? & X & X \\ 

ProReveal\cite{Jo-ProReveal}
& O & \makecell[c]{Python / TS}
& O & O & O & X & X & -
& $N$ & O & \makecell[c]{RR / FIFO} & O & O \\

DataSite\cite{Cui-DataSite}
& X & JS / JS
& O & O & O & $k$-means, DBSCAN & X & Regression 
& $N$ & X & Concurrent & X & O \\

ProBGP\cite{Ulmer-ProBGP}
& X & ? / ?
& O & O & X & X & X & Shortest path
& 1 & X & - & - & X \\

ProSteer\cite{Steering-by-Example-2022-Hografer-HansJorg} %
& O & Python / TS 
& O & O & X & X & X & -
& 1 & O & - & - & X \\ 

ProSample\cite{Prosample-Tailorable-Sampling-for-PVA-2024-Hografer}
& O & Python / TS 
& O & X & X & X & X & Sampling 
& 2 & X & Concurrent & X & X \\ 

ProInterest\cite{ProInterest-Combining-Degree-of-Interest-and-PVA-2023-Hografer}
& O & Python / TS
& O & O & X & X & X & -
& 1 & O & -  & - & X \\ 

PEViz\cite{Zhang-PEViz}
& X & ? / ?
& O & O & O & X & X & -
& $N$ & O & ? & X & O \\

MolSieve\cite{MolSieve-PVA-for-molecular-dynamics-simulation-2024-Hnaatyshyn}
&O & \makecell[c]{Python / JS}
& O & O & X & OPTICS & X & -
& $N$ & O & \makecell[c]{Concurrent}  & O & O \\

PIVOT\cite{PIVOT-viz-oriented-progressive-timeseries-transformation-2025-Chen-Yunhai}
& O& C++ / TS
& O & O & O & X & X & -
& 1 & X & - & - & X \\

ParcoursVis\cite{parcoursvis-visualization-of-electronic-health-record-sequence-2025-Ambre-Fekete}
& O & \makecell[c]{Python + C++ / JS}
& O & O & O & X & X & -
& 1 & X & - & - & X \\

\midrule
\multicolumn{12}{l}{Toolkits / Grammars} \\
\midrule

ProgressiVis\cite{Fekete-ProgressiVis}
& O & Python / -
& O & O & O & $k$-means & PCA, $t$-SNE & Quartile
& $N$ & O & RR  & X & O \\

P5\cite{kwan-liu-P5}
& O& - / JS
& O & O & X & X & X & -
& $N$ & O & RR  & X & O \\

ProVega\cite{provega-grammar-ease-prototyping-2026-filosa-marco}
& O& - / JS
& O & O & O & X & $t$-SNE & -
& 1 & O & - & - & O \\

\textbf{ProGram (ours)}
& O & Python / JS
& O & O & O & $k$-means & PCA, $t$-SNE & Quartile 
& $N$ & O & \makecell[c]{Chunked RR}  & O & O \\

\bottomrule
\end{tabular}%
}
\caption{Survey of progressive visual analytics systems, toolkits, and grammars.
\textit{Architecture} records the implementation languages of the server- and client-side components when identifiable.
\textit{Analytical Coverage} indicates whether the system supports selection/filtering, count/sum, min/max, clustering, and projection; a task is marked `O' only when its result is visible to the user, so operations used internally for another computation are not counted (e.g., ProBGP~\cite{Ulmer-ProBGP} applies DBSCAN to approximate centroids without exposing it).
\textit{Progressive Control by the User} describes how far each system lets the user adapt an ongoing computation to what they need at that moment, covering concurrent queries, pausing and resuming, scheduling, query prioritization, and ad-hoc querying.
Concurrent queries count the queries that can progress independently rather than the views that display the same computation; $N$ denotes systems in which users can add visualizations and run them concurrently.
`O' and `X' indicate that a property is supported or not supported, respectively.
`?' indicates that a property could not be verified, for example, when a paper implies a server-side component but does not report its implementation language.
`-' indicates that a property does not apply, such as scheduling or query prioritization in a single-query system.}
\label{table:pva_dev_survey}
\vspace{-3mm}
\end{table*}

We survey prior PVA systems, toolkits, and grammars to understand how they can be adopted in practice.
Rather than characterizing design decisions as prior surveys did~\cite{The-human-user-in-PVA-2019-Micallef, Angelini-A-review-and-characterization-of-PVA, A-Survey-on-Progressive-Visualization}, we ask which tools are currently available to developers, which analytical tasks and forms of progressive control they support, and which implementation burdens arise when extending them.

\subsection{Selection Criteria}
We take the 48 articles surveyed by Ulmer et al.~\cite{A-Survey-on-Progressive-Visualization} as our baseline set.
Since our focus is on systems that developers can readily use, we first consider the 14 articles that provide publicly available source code, and retain 9 of them: 7 application-oriented PVA systems with user interfaces and 2 PVA toolkits or grammars.
The remaining 5 are excluded as theoretical or algorithmic contributions without an interactive analysis environment.
From the 34 articles without available code, we additionally include 7 that describe their architecture and progressive execution behavior in sufficient detail.
Finally, to cover work published after the survey, we search Google Scholar for ``progressive visualization'' and ``progressive visual analytics,'' which yields 5 more articles, resulting in a final collection of 21.

\subsection{Analysis Dimensions}
We define four implementation-oriented dimensions from the developer's perspective.
We first record whether each system is \textit{open-source}, since this determines whether its code can be reused or extended at all.
We then examine its \textit{Architecture}, that is, whether computation is client-only or split across server and client, together with the implementation languages when identifiable.
Under \textit{Analytical Coverage}, we categorize the analytical tasks each system supports (e.g., filtering or clustering), which indicates how far an existing system meets a developer's needs before custom code becomes necessary.
Finally, \textit{Progressive Control by the User} asks how much of the control that PVA is expected to offer, such as running several queries concurrently, pausing and resuming computation, or prioritizing one query over another, is already available in existing systems, and therefore how much a developer can expect to inherit rather than build.

\subsection{Implementation Gaps}

\cref{table:pva_dev_survey} summarizes our survey.
Of the 21 articles, 18 present a PVA system built for a specific analytical task with a user interface, while one contributes a toolkit~\cite{Fekete-ProgressiVis} and two others a grammar~\cite{kwan-liu-P5, provega-grammar-ease-prototyping-2026-filosa-marco}; we list the latter three separately, as they are general frameworks for building PVA systems rather than systems themselves.
The systems are largely uniform in shape: all 18 provide a dedicated client, and every one whose implementation language we could identify (13 of 18) uses a web-based interface written in JavaScript or TypeScript, while 16 of 18 articles describe a dedicated server, most often implemented in Python (9 of the 12 we could identify).
Yet only 9 of the 18 systems make their source code available.
From this survey, we derive three implementation gaps, each grounded in one of our analysis dimensions, that inform the design of \ours.

\textbf{G1. Fragmented Execution Logic.}
Progressive computation pipelines are often split across two runtimes.
Most surveyed systems pair a Python server with a JavaScript or TypeScript client, an arrangement that is well-motivated, since these systems assume data sizes that make client-side processing impractical, but that leaves developers to bridge two languages and two notions of state.
Progressive execution can make this bridge more demanding as intermediate results must cross the boundary repeatedly rather than once.
The split is therefore not a design flaw to be removed but a constraint to be absorbed, and at present it is absorbed by developers, who rewrite the same coordination logic for each system.

\textbf{G2. Lack of Reusable Analytical Operations.}
Analytical coverage is narrow and repetitive across the surveyed systems.
Selection and filtering appear in all 18 systems and simple aggregates such as count and sum in 14, whereas minimum and maximum appear in 6, clustering in 4, and projection in only 3.
The same handful of operations is thus reimplemented system after system, each time in an application-specific form and tied to a particular data model, task, and update schedule.
The three reusable frameworks do provide operations as shared functionality, but none of them targets the architecture that \textbf{G1} identifies as dominant;
ProgressiVis places its operations on the Python side and assumes a notebook front end, P5 assumes data processing happens in the browser, and ProVega reuses Vega-Lite transformations and permits external analytical modules but does not itself provide a server-side computation runtime.

\textbf{G3. Lack of Reusable Progressive Control.}
Progressive control is realized ad hoc, in forms that do not transfer between systems.
Only 10 of the 18 systems run multiple queries that progress independently, and among these, scheduling is heterogeneous where it is described at all, spanning round-robin, first-in-first-out, and thread-pool strategies, with 5 of the 10 leaving it undocumented.
Pausing, resuming, and prioritization follow the same course: each is offered by several systems and not others, and in every case as a behavior wired into the system rather than a capability a developer can adopt.
Prioritization in particular appears in only 3 of the 10 concurrent systems, and elsewhere only in weaker task-level variants in which selecting a pattern or a region of interest causes the system to load or explore that portion first~\cite{PPMT-Progressive-sequential-pattern-mining-Raveneau, Steering-by-Example-2022-Hografer-HansJorg, Prosample-Tailorable-Sampling-for-PVA-2024-Hografer, ProInterest-Combining-Degree-of-Interest-and-PVA-2023-Hografer}.
The three reusable frameworks close only part of this gap.
ProgressiVis and P5 both support concurrent queries, pausing and resuming, and round-robin scheduling, but neither exposes prioritization as an explicit capability: ProgressiVis reschedules modules affected by an interaction under a tighter quantum, which improves responsiveness but without letting users assign priorities among concurrent queries, and P5 does not address prioritization at all.
ProVega provides per-workflow controls such as pause and play but does not expose scheduling or prioritization among concurrent queries.


\section{The ProGram Grammar}
\label{section:objectives}

\subsection{Design Rationale}
We aim to lower the barrier for developers building progressive computation pipelines.
The gaps identified above translate into a set of design requirements, which we describe below and which together informed the design of \ours.

\textbf{DR1. Driving Computation from the Client.}
To address the fragmentation of execution logic in distributed architectures (\textbf{G1}), we adopt a client-driven design in which the client takes the initiative in controlling progressive computation pipelines, while the engine functions as a computation engine.
Placing control on the client keeps user input, query specification, and visual output in one runtime, so that mapping an interaction to a query and back to a result no longer crosses the architecture boundary.
The engine, in turn, receives specifications rather than application logic, and its implementation stays independent of the analyses it serves.

\textbf{DR2. Providing Reusable Analytical Operations.}
To address the reimplementation of basic analytical operations (\textbf{G2}), we expose common operations as reusable components of the grammar rather than leaving them embedded in application-specific code.
We organize the computation specification into five primitives: \textit{Chunking} defines how a query advances, either by loading a batch of records or by running a bounded number of iterations, \textit{Filtering} selects which loaded data enter the computation, \textit{Transforming} derives refined data from the filtered one, 
\textit{Binning} progressively updates the binning state from the transformed data. \textit{Reducing} turns that state into a transmissible result.

\textbf{DR3. Providing Progressive Control Throughout the Analysis.}
To address the ad hoc and non-transferable realization of progressive control (\textbf{G3}), we make control part of the specification rather than of each system's implementation.
Progressive control enters the specification at three points: it is \textit{negotiated} before execution, \textit{invoked explicitly} during execution, and \textit{triggered implicitly} by ordinary interactions with the updating view.
Before execution, a query may optionally perform a \textit{handshake}, in which the client and server negotiate execution parameters such as chunk size or expected response time.
During execution, the grammar exposes explicit operations, such as pausing, resuming, and stopping queries, allowing users to adjust the computation as their interests shift, as resources become scarce, or as partial results stabilize.
Interactions that users already perform on the updating view can also steer the computation, without the user ever addressing the query itself.
Prioritization is one such case: it makes the details-on-demand step of the visual information-seeking mantra~\cite{information-seeking-mantra-the-eyes-have-it-SHNEIDERMAN} available even while other computations are still running.
When users select a region of interest and drill down to create a new visualization, the query for that visualization is given priority, so details for the selection arrive quickly while the remaining computations continue at a lower priority.

\textbf{DR4. Sharing Computation across Queries.}
The preceding requirements let a client run many queries at once, and those queries frequently touch the same data; for example, a dashboard typically shows several views over one table.
Progressive systems we surveyed in \cref{table:pva_dev_survey} have little means of exploiting this: each query is specified in isolation, with no way to state that it reads the same source as another or builds on what another has already computed.
We therefore make relationships between queries expressible, turning them into opportunities the engine can exploit for efficiency.
We introduce two such relationships: \textit{lockstepping} and \textit{forking}.
Lockstepping processes queries over the same data source in step, so that a chunk is read once and delivered to every query that needs it.
Forking lets a query derive from a running one and reuse its intermediate state instead of starting from raw records; for example, if a query is already maintaining the binning state for a 2D histogram, a 1D histogram over one of its dimensions can be forked from that state and marginalized from the existing bins, producing its first partial result without repeating work the engine has already done.

\subsection{Computing a Single Progressive Query}

The minimal executable unit of \ours is a single \emph{query}, which specifies the progressive operations and their parameters required to serve a single visualization (\cref{code:progressive_query_example}).
\ours employs \textbf{chunking} to achieve progressive execution, splitting a computation either by data (batch-based) or by iterations (iteration-based).
Each execution chunk then passes through a pipeline of progressive operations; we support four types of operations: \textbf{filtering}, \textbf{transforming}, \textbf{binning}, and \textbf{reducing}.
Before committing to a query, the client may optionally perform a \textbf{handshake}, in which the engine evaluates the intended computation over a limited portion of the data and returns a preview together with execution statistics, allowing the client to adjust the execution parameters in advance.

In implementation, a query is written as a JSON object and sent to a server-side  engine through a client API.
A client connects with \code{pg = ProGram("address")} and submits a query specification $Q$ with \code{pg.send($Q$)}.
Progressive results are streamed back over a socket to a handler registered with \code{pg.on($Q$, (result) => render(result))}, which the client uses to re-render its visualization on each update.

\begin{listing}[t]
\centering

\begin{tcolorbox}[
    enhanced,
    width=\linewidth,
    colback=black!1,
    colframe=black!18,
    boxrule=0.4pt,
    arc=1.5mm,
    title={ProGram Query Example},
    colbacktitle=black!7,
    coltitle=black!85,
    fonttitle=\footnotesize\sffamily\bfseries,
    left=1.5mm,
    right=1.5mm,
    top=1.3mm,
    bottom=1.3mm,
]

\scriptsize
\ttfamily

\schematag{queryColor}{ProGramQuery} \{

\vspace{0.5mm}

\hspace*{3mm}\qfield{from}:
    \qlit{"movie\_dataset"}, \\
\hspace*{3mm}\qfield{chunkType}:
    \qlit{"batch"}, \qfield{batchSize}: \qlit{1000}, \\
\hspace*{3mm}\qfield{responseInterval}:
    \qlit{1000}, \\
\hspace*{3mm}\qfield{ops}: [\{ \\
\hspace*{10mm}\qfield{type}:
    \schematag{transformingColor}{"transform"}, \\
\hspace*{10mm}\qfield{name}:
    \qlit{"profit"}, \qfield{expr}:
    \qlit{"revenue - production\_cost"} \\
\hspace*{6mm}\},
\{ \\
\hspace*{10mm}\qfield{type}:
    \schematag{filteringColor}{"filter"}, \\
\hspace*{10mm}\qfield{pred}:
    \qlit{"profit > 0"} \\
\hspace*{6mm}\}, 
\{ \\
\hspace*{10mm}\qfield{type}:
    \schematag{binningColor}{"bin"}, \\
\hspace*{10mm}\qfield{id}:
    \qlit{"audience\_by\_genre"}, \\
\hspace*{10mm}\qfield{keys}: [\{ \qfield{field}: \qlit{"genre"}\}], \\
\hspace*{10mm}\qfield{values}: [\{ \\
\hspace*{16mm}\qfield{name}:
    \qlit{"admission"}, \qfield{field}:
    \qlit{"audience"}, \\
\hspace*{16mm}\qfield{op}:
    \qlit{"sum"}, \qfield{confidence}:
    \qlit{0.95} \\
\hspace*{13mm}\}, \\[0.3mm]
\hspace*{13mm}\{ \\
\hspace*{16mm}\qfield{name}:
    \qlit{"movie\_count"}, \qfield{op}: \qlit{"count"} \\
\hspace*{13mm}\}] \\
\hspace*{6mm}\},
\{ \\
\hspace*{10mm}\qfield{type}:
    \schematag{reducingColor}{"reduce"}, \\
\hspace*{10mm}\qfield{method}:
    \qlit{"top\_k"}, \\
\hspace*{10mm}\qfield{params}: \{ \\
\hspace*{13mm}\qfield{bins}:
    [\qlit{"audience\_by\_genre"}], \qfield{k}: \qtype{5}, \\
\hspace*{13mm}\qfield{value}:
    \qlit{"admission / movie\_count"}, \\
\hspace*{13mm}\qfield{order}:
    \qlit{"descending"} \\
\hspace*{10mm}\} \\
\hspace*{6mm}\}] \\[0.3mm]
\}
\end{tcolorbox}
\vspace{-3mm}
\caption{An example of a \ours query that computes the mean audience size of profitable movies by genre.}
\vspace{-4mm}
\label{code:progressive_query_example}
\end{listing}


\begin{listing}[t]
\centering

\begin{tcolorbox}[
    enhanced,
    width=\linewidth,
    colback=black!1,
    colframe=black!18,
    boxrule=0.4pt,
    arc=1.5mm,
    title={ProGram Query Grammar},
    colbacktitle=black!7,
    coltitle=black!85,
    fonttitle=\footnotesize\sffamily\bfseries,
    left=1.5mm,
    right=1.5mm,
    top=1.3mm,
    bottom=1.3mm,
]

\scriptsize
\ttfamily

\schematag{queryColor}{ProGramQuery}
\ \textcolor{black!55}{::=}\ 
\textcolor{typeColor}{Query}
\ \textcolor{black!45}{|}\ 
\textcolor{typeColor}{Handshake}
\\
\hspace*{26mm}
\textcolor{black!45}{|}\ 
\textcolor{typeColor}{Control}
\textcolor{black!45}{|}\ 
\textcolor{typeColor}{Configure}
\vspace{1.2mm}

\textcolor{typeColor}{Query} \{

\vspace{0.4mm}

\hspace*{4mm}\qfield{id}?:
    \textcolor{typeColor}{String}, \qfield{from}: \textcolor{typeColor}{String}, \\
\hspace*{4mm}\qfield{chunkType}:
    \qlit{"batch"} \textcolor{black!45}{|} \qlit{"iter"} \textcolor{black!45}{|} \qlit{"both"}, \\
\hspace*{4mm}\qfield{batchSize}?:
    \textcolor{typeColor}{Number}, \qfield{iterSize}?: \textcolor{typeColor}{Number}, \\
\hspace*{4mm}\qfield{responseInterval}:
    \textcolor{typeColor}{Number}, \\
\hspace*{4mm}\qfield{priority}?:
    \textcolor{typeColor}{Number}, \\
\hspace*{4mm}\qfield{lockstep}?:
    \textcolor{typeColor}{String}, \\
\hspace*{4mm}\qfield{ops}:
    \textcolor{typeColor}{Operation[]}, \\
\hspace*{4mm}\qfield{execution}?: \qlit{"cpu"} | \qlit{"gpu"}

\vspace{0.3mm}

\}

\vspace{1.2mm}

\textcolor{typeColor}{Handshake} \{

\vspace{0.4mm}

\hspace*{4mm}\qfield{from}:
    \textcolor{typeColor}{String}, \\
\hspace*{4mm}\qfield{type}:
    \qlit{"handshake"}, \\
\hspace*{4mm}\qfield{sampleSize}?:
    \textcolor{typeColor}{Number}, \qfield{maxTime}?: \textcolor{typeColor}{Number}, \\ 
\hspace*{4mm}\qfield{ops}:
    \textcolor{typeColor}{Operation[]}

\vspace{0.3mm}

\}

\vspace{0.8mm}


\vspace{0.8mm}


\end{tcolorbox}
\vspace{-3mm}
\caption{Computation and handshake queries}
\label{code:program_query_grammar}
\vspace{-4mm}
\end{listing}


\begin{listing}[t]
\centering

\begin{tcolorbox}[
    enhanced,
    width=\linewidth,
    colback=black!1,
    colframe=black!18,
    boxrule=0.4pt,
    arc=1.5mm,
    title={ProGram Operation Grammar},
    colbacktitle=black!7,
    coltitle=black!85,
    fonttitle=\footnotesize\sffamily\bfseries,
    left=1.5mm,
    right=1.5mm,
    top=1.3mm,
    bottom=1.3mm,
]

\scriptsize
\ttfamily

\textcolor{typeColor}{Operation}
\ \textcolor{black!55}{::=}\ 
\schematag{transformingColor}{Transform}
\ \textcolor{black!45}{|}\ 
\schematag{filteringColor}{Filter}
\textcolor{black!45}{|}\ 
\schematag{binningColor}{Bin}
\ \textcolor{black!45}{|}\ 
\schematag{reducingColor}{Reduce}

\vspace{1.1mm}

\schematag{transformingColor}{Transform} \{ \\
\hspace*{4mm}\qfield{type}: \qlit{"transform"}, \\
\hspace*{4mm}\qfield{name}: \textcolor{typeColor}{String}, \\
\hspace*{4mm}\qfield{expr}?: \textcolor{typeColor}{String}, \\
\hspace*{4mm}\qfield{model}?: "pca" | "tsne" | "kmeans" | "online\_kmeans",\\
\hspace*{4mm}\qfield{params}?: {\{...\}} // Transformer-specific parameters \\
\}
\vspace{1.0mm}

\schematag{filteringColor}{Filter} \{ \\
\hspace*{4mm}\qfield{type}: \qlit{"filter"}, \\
\hspace*{4mm}\qfield{pred}: \textcolor{typeColor}{String} \\
\}

\vspace{1.0mm}

\schematag{binningColor}{Bin} \{ \\
\hspace*{4mm}\qfield{type}: \qlit{"bin"}, \\
\hspace*{4mm}\qfield{id}: \textcolor{typeColor}{String}, \\
\hspace*{4mm}\qfield{keys}: \textcolor{typeColor}{KeyDef[]}, \qfield{values}: \textcolor{typeColor}{ValueDef[]} \\
\}

\vspace{0.5mm}

\textcolor{typeColor}{KeyDef} \{ 
\qfield{field}: \textcolor{typeColor}{String}, \qfield{bin}?: \textcolor{typeColor}{Number} 
\}

\vspace{0.5mm}

\textcolor{typeColor}{ValueDef} \{ \\
\hspace*{4mm}\qfield{name}?: \textcolor{typeColor}{String}, \\
\hspace*{4mm}\qfield{field}?: \textcolor{typeColor}{String | String[]}, \\
\hspace*{4mm}\qfield{op}: "count" | "sum" | "min" | "max" | "sqsum" | \\ \hspace*{10mm}"quartile", \\
\hspace*{4mm}\qfield{confidence}?: \textcolor{typeColor}{Number} \\
\}


\vspace{1.0mm}

\schematag{reducingColor}{Reduce} \{ \\
\hspace*{4mm}\qfield{type}: \qlit{"reduce"}, \\
\hspace*{4mm}\qfield{method}: "top\_k" | "filter" | \\ \hspace*{14mm} "heatmap\_grid" | "contour",\\ 
\hspace*{4mm}\qfield{params}: {\{...\}} // Reducer-specific parameters \\
\}
\end{tcolorbox}
\vspace{-3mm}
\caption{\ours operation grammar}
\label{code:program_stage_grammar}
\vspace{-3mm}
\end{listing}


\begin{listing}[t]
\centering

\begin{tcolorbox}[
    enhanced,
    width=\linewidth,
    colback=black!1,
    colframe=black!18,
    boxrule=0.4pt,
    arc=1.5mm,
    title={ProGram Interaction Grammar},
    colbacktitle=black!7,
    coltitle=black!85,
    fonttitle=\footnotesize\sffamily\bfseries,
    left=1.5mm,
    right=1.5mm,
    top=1.3mm,
    bottom=1.3mm,
]

\scriptsize
\ttfamily



\textcolor{typeColor}{Control} \{ \\
\hspace*{4mm}\qfield{action}:
    \qlit{"pause"}
    \textcolor{black!45}{ | }
    \qlit{"resume"}
    \textcolor{black!45}{ | }
    \qlit{"stop"}, \\
\hspace*{4mm}\qfield{targetQueryId}: \textcolor{typeColor}{String} \\
\}

\vspace{0.8mm}

\textcolor{typeColor}{Configure} \{ \\
\hspace*{4mm}\qfield{action}: \qlit{"configure"}, \\
\hspace*{4mm}\qfield{targetQueryId}: \textcolor{typeColor}{String}, \\
\hspace*{4mm}\qfield{config}: \textcolor{typeColor}{Config} \\
\}

\vspace{0.8mm}

\textcolor{typeColor}{Config} \{ \\
\hspace*{4mm}\qfield{batchSize}?: \textcolor{typeColor}{Number}, \\
\hspace*{4mm}\qfield{iterSize}?: \textcolor{typeColor}{Number}, \\
\hspace*{4mm}\qfield{reduce}?: \textcolor{typeColor}{Reduce}, \\
\hspace*{4mm}\qfield{schedule}?: \{ \qfield{priority}: \textcolor{typeColor}{Number}, \qfield{duration}?: \textcolor{typeColor}{Number} \}, \\
\hspace*{4mm}\qfield{lockstep}?: \textcolor{typeColor}{String | null} \\
\}

\vspace{0.8mm}


\end{tcolorbox}
\vspace{-3mm}
\caption{Control and configure queries}
\label{code:program_interaction_grammar}
\vspace{-1mm}
\end{listing}

\paragraph{Handshaking}
Before query execution, the client may optionally perform a handshake to preview the result of a computation and adjust query parameters (\code{Handshake} in \cref{code:program_query_grammar}, \textbf{DR3}). 
Prior work notes that users may explore an analytic parameter space by comparing multiple executions under different settings~\cite{The-human-user-in-PVA-2019-Micallef}. 
A naive approach is to repeatedly start a query with a chosen set of parameters, inspect its response time, and manually terminate the executions that appear unsuitable. 
A handshake provides a bounded alternative to this trial-and-error process.

The client submits the intended computation along with constraints such as a sample size or a time budget.
The engine evaluates the specification over a limited portion of the data and reports execution statistics, including the number of records processed, the elapsed time, and an estimate of the cost of a full execution under current server conditions.
The client can then adjust parameters such as the chunk size, the binning granularity, or the reduction strategy before committing to the query.
The engine also returns the result of the operations, which the client can use to serve an initial visualization.

\paragraph{Chunking} 

\ours realizes progressiveness by decomposing a query into a sequence of execution chunks.
Each chunk performs a bounded amount of work and updates the intermediate state of the query, from which a new partial result is produced. 
We support three bases for defining these chunks: batch-based chunking (\code{chunkType: "batch"}), iteration-based chunking (\code{chunkType: "iter"}), and using both schemes (\code{chunkType: "both"}) motivated by a prior survey~\cite{A-Survey-on-Progressive-Visualization}. 
\textbf{Batch-based chunking}, the dominant scheme in the prior survey (37 of 48 articles), forms each execution chunk from a fixed number of records and passes it through the pipeline.

In \textbf{iteration-based chunking}, in contrast, each execution chunk consists of a fixed number of iterations of a transformer.
For example, with a $k$-means clustering transformer, an iteration size of three performs three clustering iterations over all points.
Note that in iteration-based chunking, the entire dataset feeds into every iteration, which assumes that at least one iteration can complete within the given response interval.

In \textbf{hybrid chunking}, an execution chunk is defined by both parameters at once: each chunk admits a fixed number of new records and then runs a fixed number of iterations over the data accumulated so far.
This scheme fits transformers whose result must be refined as data arrive, such as online $k$-means clustering, where every chunk assigns the incoming records to the current centroids and then updates the centroids over the accumulated points for the specified number of iterations.

To meet the given response interval, the engine tracks how long the last chunk took to process.
Before starting the next chunk, it checks whether adding this estimate to the time since the previous response would exceed the interval.
For example, suppose a query sets a response interval of one second and the engine estimates 240 ms per chunk.
Once 760 ms have passed since the previous response, processing another chunk would exceed the interval, so the engine returns the current partial result first and processes the chunk afterwards.

The chunk size determines the update granularity of a query, introducing a trade-off between latency guarantee the engine can provide and throughput.
A small chunk size allows the engine to return a response close to the specified interval, at the cost of throughput, as execution-time estimation and scheduling interrupt the computation more frequently.
A large chunk size coarsens this granularity, since the elapsed time is examined only at chunk boundaries.
The engine may then respond substantially before the interval elapses, leaving part of the budget unused, or spend longer than the interval on a single chunk and delay the response beyond it.

\paragraph{Filtering and Transforming}
A filter consists of one or more predicates and selects the subset of the incoming data that is passed to the remaining operations (\code{Filter} in \cref{code:program_stage_grammar}).
For example, the query in \cref{code:progressive_query_example} retains only the movies whose profit is greater than zero.

A transforming operation consumes one or more input fields and produces derived fields (\code{Transform} in \cref{code:program_stage_grammar}).
\ours supports two types of transforming operations.
An expression-based transformer is specified by a \code{name} and an \code{expr}; the engine evaluates \code{expr} over each record and appends the result as a derived field named \code{name}.
For example, the query in \cref{code:progressive_query_example} derives a profit field by subtracting the production cost from the revenue of each record.
A model-based transformer instead applies a predefined model, such as dimensionality reduction or clustering. 
Filtering and transforming operations can be interleaved: a query may derive a net profit field (transforming), retain only the movies whose net profit is greater than zero (filtering), and then cluster the retained movies by their attributes (transforming).

\paragraph{Binning}
A binning operation incorporates the outputs of filtering and transforming into a persistent server-side binning state maintained for each query (\code{Bin} in \cref{code:program_stage_grammar}).
It is specified by a set of keys and a set of values.
Keys determine how the incoming records are partitioned, for example by a categorical attribute, a numerical interval, or a combination of multiple dimensions.
Values define the statistics maintained for each key, such as count, sum, squared sum, minimum, maximum, and quartile.
When the uncertainty of an aggregate can be estimated, as for count, sum, and squared sum, the specification may additionally request a confidence interval for the maintained value.
For example, the \code{bin} operation in \cref{code:progressive_query_example} partitions the movies by genre and maintains the sum of the audience and the count of movies, with a confidence interval at the 0.95 confidence level.

The binning state is retained across execution chunks and incrementally updated for the bins affected by newly processed records.
The granularity of the keys therefore introduces a trade-off between memory consumption and the analytical detail preserved for subsequent operations.
Fine-grained or high-dimensional keys produce more bins and require additional memory, but they preserve information that coarser bins discard irrecoverably.
We thus recommend maintaining bins as fine as memory permits, since a state computed once can be reused when the user later requests a coarser view, whereas a finer state must be recomputed from the beginning.
A fine-grained state, however, increases the amount of data transmitted to the client; this cost can be reduced by a subsequent reducing operation, for which each binning state is assigned an identifier that later reducers can reference.
For example, the query in \cref{code:progressive_query_example} assigns the identifier \code{audience\_by\_genre} to the binning state, which the reducer then references.

\paragraph{Reducing}
A reducing operation derives a partial result from one or more binning states and returns it to the client (\code{Reduce} in \cref{code:program_stage_grammar}).
A reducer mitigates the transmission overhead of a fine-grained state when only part of the maintained information is required for the visualization.
\ours supports four types of reducers: 
(1) aggregating finer bins into coarser ones, for example merging $1{,}024 \times 1{,}024$ bins into a $64 \times 64$ grid;
(2) filtering a subset of bins, for example keeping only the bins within a given range; (3) selecting a top-$k$ result, as in \cref{code:progressive_query_example}; (4) converting a binning state into contours, which yields, for each level, the sequence of coordinates tracing the corresponding contour line.

\subsection{Computing Multiple Progressive Queries}
A progressive system often executes multiple progressive queries concurrently, for example when it serves a dashboard composed of several views, which was the case for 10 of the 18 systems in our survey.
We therefore extend the single-query execution model by scheduling queries and coordinating the states shared among them.

\begin{figure}[t] 
 \centering 
 \includegraphics[width=\columnwidth]{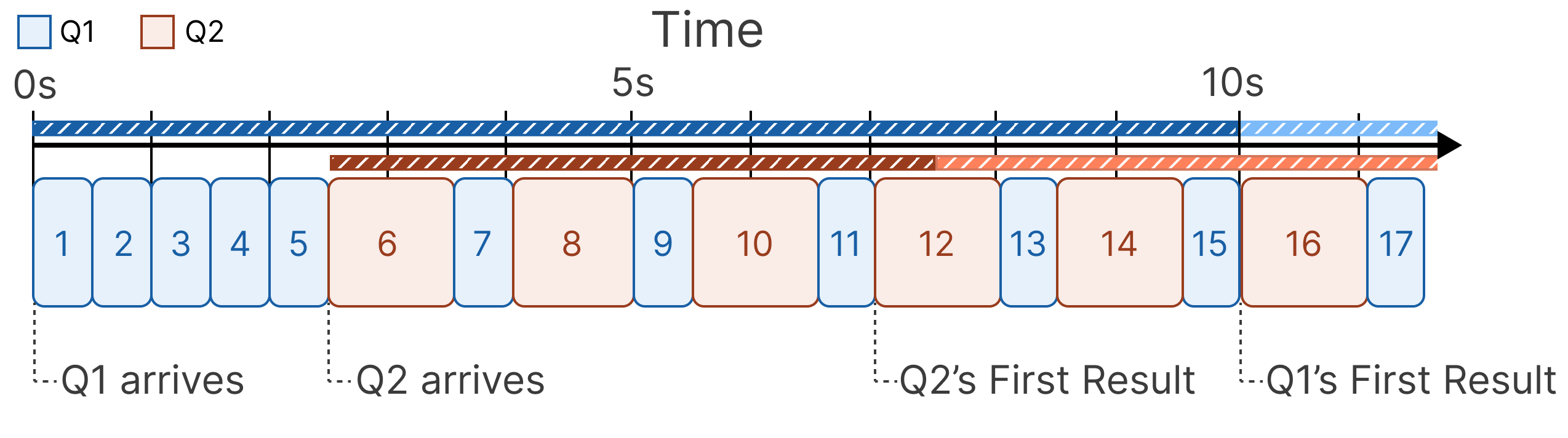}
 \vspace{-3mm}
 \caption{Example of chunked round-robin scheduling. Q1 (blue) starts at $t=0$ with a 10-second response interval and processes two chunks per second, while Q2 (red) arrives at $t=2.5$ with a 5-second response interval and processes one chunk per second. Response intervals are drawn as hatched rectangles along the time axis, with the color alternating between consecutive intervals. Once Q2 arrives, the scheduler interleaves the two queries at chunk granularity. Q2's first response interval ends at $t=7.5$, but the engine returns its partial result at $t=7$, since no further Q2 chunk could complete before the deadline.}
 \label{fig:scheduling}
 \vspace{-3mm}
\end{figure}

By default, the engine schedules queries in a \emph{chunked round-robin} manner, in which each query executes one chunk before yielding control to the scheduler.
Conventional round-robin scheduling allocates a fixed time quantum to each task, whereas our scheduling unit is a chunk of progressive computation; the duration of a turn therefore varies with the query and its chunk size.
After a chunk completes, the scheduler selects the next query in cyclic order, so that multiple queries progress without preemption within a chunk.
For example, \cref{fig:scheduling} illustrates the execution of two queries with different processing rates and response intervals.
Q1 (blue) starts first and is processed continuously until Q2 (red) arrives.
Once Q2 becomes active, the scheduler alternates between the two, assigning one chunk to each query per turn.
Each query returns its accumulated partial result according to its own response interval; if the time remaining before the next response is insufficient for another chunk, the current partial result is returned before processing that chunk.

Under concurrent execution, the chunk size affects the schedule beyond the trade-off between the latency guarantee and throughput discussed above.
Since a query enters the cycle only at a chunk boundary, a smaller chunk shortens the delay before a newly arrived query is first served, at the cost of more frequent scheduling.
The engine further assigns a priority to each query: queries with higher priorities are scheduled first, and those at the same priority follow the chunked round-robin policy.
Priorities allow the engine to serve a query issued in response to a user interaction ahead of the queries already in progress, which we discuss in \cref{sec:interaction}.

\textbf{Lockstepping} coordinates multiple batch-based queries so that the engine loads each data batch once and distributes it to every query in the group, which improves I/O efficiency (\textbf{DR4}).
Each query in the group then applies its own operations to the shared input, and queries can be lockstepped only when they draw on an identical data source and use identical chunking.
A group is formed by naming it in the \code{lockstep} field of a query, and every query that specifies the same name joins that group.
When queries join a group at different progress levels, the engine aligns them before sharing subsequent batches by pausing the queries that are further ahead.

\textbf{Forking} lets a newly submitted query reuse a binning state already maintained by an executing query instead of recomputing it from scratch (\textbf{DR4}).
The engine detects such an opportunity automatically by comparing the filtering and transforming operations, as well as the binning keys and values, of the new query with those of the executing ones.
For example, if an existing query maintains the total audience and the movie count for each genre, a new query can reuse the binning state to present a different ranking or result representation without reprocessing the records consumed so far.
A forked query starts from the progress of its source query and proceeds independently thereafter, and it may use a different reducer, scheduling priority, or chunk size, as long as its initial binning state is derivable from the existing ones.

\subsection{Involving the User during Progressive Execution}\label{sec:interaction}
\ours supports control over progressive queries through three mechanisms that connect client-side interactions to the progressive execution model: progress control, specification reconfiguration, and temporary prioritization (\textbf{DR3}).

\paragraph{Progress control}
A client may pause, resume, or terminate a running query, which was the most frequent form of control in our survey, appearing in 10 of the 18 systems.
These interactions are expressed as control queries (\code{Control} in \cref{code:program_interaction_grammar}).
Because the queries in a lockstep group consume the same data batch at each execution chunk, a pause or resume query applies to the entire group: pausing a single member would leave it behind the shared progress and break the alignment the group maintains.

\paragraph{Specification Reconfiguration}
During execution, a client may reconfigure selected components of an ongoing query without restarting it (\code{Configure} in \cref{code:program_interaction_grammar}).
Existing systems support such reconfiguration only to a limited extent, typically re-prioritization alone, whereas \ours admits a wider range: changing the reducer, adjusting the chunk size, and adding or removing lockstepping.
A reconfiguration request is not applied immediately while an execution chunk is in progress; it takes effect at the next chunk boundary.

The cost of a reconfiguration depends on which component it touches, and falls into three levels.
Changing the scheduling priority or the chunk size affects only how the query is scheduled, so it takes effect from the next chunk and leaves every computed result intact; detaching a query from a lockstep group is equally inexpensive, as the query simply proceeds on its own from the point it has reached.
Changing the reducer costs more, since the new reducer must be evaluated over the retained bins, but it still avoids reprocessing any record as long as the maintained bins contain every value the new result requires.
Consider a movie analysis in which a progressive bar chart compares genres by the mean audience of profitable movies: the binning state maintains the total audience and the movie count for each genre, from which the client can also derive the total audience per genre, or rank the genres by a different measure, without revisiting the records observed so far.
Reconfiguring filtering, transforming, or binning is the most expensive case and generally requires a restart, since a new filter admits records that the current state never counted, and a finer key cannot be recovered from bins that have already merged them.

\paragraph{Temporary Prioritization}
A query triggered by a tooltip or another transient interaction in the details-on-demand step should temporarily precede the queries already in progress, so that the user receives quick feedback on the part of the data they are currently interested in.
To support such a temporary elevation of priority, \ours allows a priority reconfiguration to carry an optional duration, after which the engine restores the previous priority of the query (\code{schedule?} in \cref{code:program_interaction_grammar}).
Setting the \code{duration} field to 5{,}000, for example, prioritizes the query for five seconds and then returns it to the value it had before.

\begin{table}[t]
\centering

\footnotesize
\begin{minipage}{\columnwidth}
\raggedright

\textbf{1. Starting a progressive query with confidence intervals, e.g., counting movies by genre with 95\% intervals for a bar chart.}

\vspace{0.5mm}

\begin{tcolorbox}[
    enhanced,
    width=\linewidth,
    colback=black!1,
    colframe=black!15,
    boxrule=0.35pt,
    arc=1mm,
    left=1.2mm,
    right=1.2mm,
    top=0.8mm,
    bottom=0.8mm
]

{\scriptsize\ttfamily
\schematag{queryColor}{ProGramQuery}\{id: "genre\_count",
  ops: [\{\\
      type: \schematag{binningColor}{"bin"},
      keys: [\{field: "genre"\}],
      values: [\{op: "count", confidence: 0.95\}]
      \}]\}
}
\end{tcolorbox}

\vspace{1.5mm}

\textbf{2. Filtering a subset of interest, e.g., selecting movies whose genre is `Action'.}

\vspace{0.5mm}

\begin{tcolorbox}[
    enhanced,
    width=\linewidth,
    colback=black!1,
    colframe=black!15,
    boxrule=0.35pt,
    arc=1mm,
    left=1.2mm,
    right=1.2mm,
    top=0.8mm,
    bottom=0.8mm
]

{\scriptsize\ttfamily
\schematag{queryColor}{ProGramQuery}\{ops: [\{type: \schematag{filteringColor}{"filter"},\\
      pred: "genre == `Action'"\}]\}
}
\end{tcolorbox}

\vspace{1.5mm}

\textbf{3. Pausing a running query, e.g., pausing the bar chart query.}

\vspace{0.5mm}

\begin{tcolorbox}[
    enhanced,
    width=\linewidth,
    colback=black!1,
    colframe=black!15,
    boxrule=0.35pt,
    arc=1mm,
    left=1.2mm,
    right=1.2mm,
    top=0.8mm,
    bottom=0.8mm
]

{\scriptsize\ttfamily
\schematag{queryColor}{ProGramQuery}\{action: "pause", targetQueryId: "genre\_count"\}
}

\end{tcolorbox}

\vspace{1.5mm}

\textbf{4. Reconfiguring a running query, e.g., prioritizing the bar chart query.}

\vspace{0.5mm}

\begin{tcolorbox}[
    enhanced,
    width=\linewidth,
    colback=black!1,
    colframe=black!15,
    boxrule=0.35pt,
    arc=1mm,
    left=1.2mm,
    right=1.2mm,
    top=0.8mm,
    bottom=0.8mm
]

{\scriptsize\ttfamily
\schematag{queryColor}{ProGramQuery}\{action: "configure",
  targetQueryId: "genre\_count", config : \{schedule: \{priority: 3\}\}\}
}
\end{tcolorbox}

\vspace{1.5mm}

\textbf{5. Providing custom features, e.g., supporting PVA-Guards.}

\vspace{0.5mm}

\begin{tcolorbox}[
    enhanced,
    width=\linewidth,
    colback=black!1,
    colframe=black!15,
    boxrule=0.35pt,
    arc=1mm,
    left=1.2mm,
    right=1.2mm,
    top=0.8mm,
    bottom=0.8mm
]

{\scriptsize\ttfamily
\begin{tabular}{@{}l@{}}
\schematag{schemaRed}{System Response}\{ ... \\
\hspace*{3mm}result: [\{ \\
\hspace*{6mm}key: \{genre: "Action"\}, \\
\hspace*{6mm}values: \{ \\
\hspace*{9mm}count: 1769313, \\
\hspace*{9mm}count\_std: 4809.3, \\
\hspace*{9mm}count\_ci: [1759886.8, 1778739.2] \\
\hspace*{6mm}\} \\
\}]\}
\end{tabular}
}

\schematag{schemaYellow}{Custom} Compute $p$-values on the client from the returned \code{values}, such as the count and its standard deviation (\code{count} and \code{count\_std}).

\end{tcolorbox}

\end{minipage}
\vspace{-1mm}
\caption{Mapping between ProReveal's features and \ours queries. Only the specifications relevant to each feature are shown.}
\label{tab:proreveal_program_mapping}
\vspace{-3mm}
\end{table}


\begin{table*}[t]
\centering

\footnotesize

\begin{minipage}[t]{0.485\textwidth}
\vspace{0pt}
\raggedright

\textbf{1. Progressive clustering with online $k$-means (\cref{fig:teaser}B-1)}

\vspace{0.5mm}

\begin{tcolorbox}[
    enhanced,
    width=\linewidth,
    colback=black!1,
    colframe=black!15,
    boxrule=0.35pt,
    arc=1mm,
    left=1.2mm,
    right=1.2mm,
    top=0.8mm,
    bottom=0.8mm
]

{\scriptsize\ttfamily
\begin{tabular}{@{}l@{}}
\schematag{queryColor}{ProGramQuery}
\{ lockstep: "kmeans\_pca",  \\ 
\hspace*{3mm}... \\
\hspace*{3mm}type:
    \schematag{transformingColor}{"transform"}, \\
\hspace*{3mm}name: "cluster",
    model: "online\_kmeans", \\
    \hspace*{3mm}params: \{fields: [...], k: 4\}, \\
... \}
\end{tabular}
}

\end{tcolorbox}

\vspace{2mm}

\textbf{2. Progressive projection with online PCA (\cref{fig:teaser}B-2)}

\vspace{0.5mm}

\begin{tcolorbox}[
    enhanced,
    width=\linewidth,
    colback=black!1,
    colframe=black!15,
    boxrule=0.35pt,
    arc=1mm,
    left=1.2mm,
    right=1.2mm,
    top=0.8mm,
    bottom=0.8mm
]

{\scriptsize\ttfamily
\begin{tabular}{@{}l@{}}
\schematag{queryColor}{ProGramQuery}
\{ lockstep: "kmeans\_pca",  \\ 
\hspace*{3mm}... \\
\hspace*{3mm}type:
    \schematag{transformingColor}{"transform"}, \\
\hspace*{3mm}name: "pca",
    model: "pca", \\
\hspace*{3mm}params: \{fields: [...], components: 2\} \\
... \}
\end{tabular}
}

\end{tcolorbox}

\vspace{2mm}

\textbf{3. Point counts per spatial cell and cluster (\cref{fig:teaser}B-3)}

\vspace{0.5mm}

\begin{tcolorbox}[
    enhanced,
    width=\linewidth,
    colback=black!1,
    colframe=black!15,
    boxrule=0.35pt,
    arc=1mm,
    left=1.2mm,
    right=1.2mm,
    top=0.8mm,
    bottom=0.8mm
]

{\scriptsize\ttfamily
\begin{tabular}{@{}l@{}}
\schematag{queryColor}{ProGramQuery}
\{ ... \\
\hspace*{3mm}type:
    \schematag{binningColor}{"bin"}, \\
\hspace*{3mm}keys: [\{field: "lon", bin: 0.0001\}, \\
\hspace*{12mm}\{field: "lat", bin: 0.0001\},
    \{field: "cluster"\}], \\
\hspace*{3mm}values: [\{op: "count"\}] \\
... \}
\end{tabular}
}

\end{tcolorbox}

\end{minipage}
\hfill
%
\begin{minipage}[t]{0.485\textwidth}
\vspace{0pt}
\raggedright

\textbf{4. Trip distance sums and counts per spatial cell and cluster (\cref{fig:teaser}B-4)}

\vspace{0.5mm}

\begin{tcolorbox}[
    enhanced,
    width=\linewidth,
    colback=black!1,
    colframe=black!15,
    boxrule=0.35pt,
    arc=1mm,
    left=1.2mm,
    right=1.2mm,
    top=0.8mm,
    bottom=0.8mm
]

{\scriptsize\ttfamily
\begin{tabular}{@{}l@{}}
\schematag{queryColor}{ProGramQuery}
\{ ... \\ 
\hspace*{3mm}type:
    \schematag{binningColor}{"bin"}, \\
\hspace*{3mm}id: "pca\_cluster\_stats", \\
\hspace*{3mm}keys: [
    \{field: "pca\_0", bin: 0.01\},
    \{field: "pca\_1", \\
\hspace*{11mm}bin: 0.01\},
    \{field: "cluster"\}], \\
\hspace*{3mm}values: [
    \{op: "sum", field: "trip\_distance"\}, \\
\hspace*{13mm}\{op: "count"\}], \\
... \}
\end{tabular}
}

\end{tcolorbox}

\vspace{1.5mm}

\textbf{5. Computing the statistics of a brushed region (\cref{fig:teaser}B-1, B-2 $\rightarrow$ B-4)}

\vspace{0.5mm}

\begin{tcolorbox}[
    enhanced,
    width=\linewidth,
    colback=black!1,
    colframe=black!15,
    boxrule=0.35pt,
    arc=1mm,
    left=1.2mm,
    right=1.2mm,
    top=0.8mm,
    bottom=0.8mm
]

{\scriptsize\ttfamily
\begin{tabular}{@{}l@{}}
\schematag{queryColor}{ProGramQuery}
\{action: "configure", \\
\hspace*{3mm}config: \{ reduce: \{ \\
\hspace*{7mm}type:
    \schematag{reducingColor}{"reduce"}, \\
\hspace*{7mm}method: "filter", \\
\hspace*{7mm}params: \{bins: ["pca\_cluster\_stats"], \\ 
\hspace*{20mm}pred:
    "pca\_0 >= -1.0 \&\& pca\_0 <= 1.0 \&\& \\
\hspace*{29mm}pca\_1 >= -0.5 \&\&
    pca\_1 <= 0.5 \&\& \\
\hspace*{29mm}(cluster == 1 || cluster == 2)"\} \\
\}\}\}
\end{tabular}
}

\end{tcolorbox}

\vspace{1.5mm}

\textbf{6. Difference view with Cohen's $d$
(\cref{fig:teaser}B-5)}

\vspace{0.5mm}

\begin{tcolorbox}[
    enhanced,
    width=\linewidth,
    colback=black!1,
    colframe=black!15,
    boxrule=0.35pt,
    arc=1mm,
    left=1.2mm,
    right=1.2mm,
    top=0.8mm,
    bottom=0.8mm
]

\schematag{schemaYellow}{Custom} Calculate Cohen's $d$ on the client using the
\code{sum}, \code{sqsum}, and \code{count} statistics of the
selected and non-selected points.

\end{tcolorbox}

\end{minipage}

\caption{Mapping between DimXplorer's features and \ours queries. Only the specifications relevant to each feature are shown.}

\label{tab:dimxplorer_program_mapping}

\end{table*}

\subsection{Engine Implementation}
\ours consists of a client library and an engine.
The client library is implemented in JavaScript and supports authoring \ours specifications, along with the WebSocket communication layer that submits queries and parses the streamed partial results into client-side data structures.
We implement the engine in Python.
The current engine only supports CSV datasets, as our goal is to design the language rather than a full-fledged data backend, and extending it to other data formats such as columnar or database-backed sources is left as future work.

To improve execution efficiency, the engine provides two execution layers: a CPU layer and a CuPy-based GPU layer.
Upon receiving a query, the engine routes it to the GPU layer if three conditions hold: (1) the client has not explicitly requested CPU execution, which may be preferable for small workloads where dispatch and data-transfer overhead outweigh the benefit of acceleration; (2) the query uses batch-based chunking; and (3) every aggregation can be expressed as a partial summary that merges associatively and commutatively.
The third condition covers count, sum, minimum, and maximum, which map directly to segmented reduction primitives in CuPy.
A histogram that progressively maintains the sum of values over fixed-size bins, for instance, satisfies all three and thus runs on the GPU.
Queries that do not satisfy these conditions fall back to the CPU layer.

The current GPU layer is limited to batch-based queries with mergeable aggregations, although other workloads could also benefit from GPU acceleration.
For example, iteration-based computations such as $t$-SNE could execute their iterative updates on the GPU while maintaining their intermediate state in GPU memory.
We leave support for these GPU executions as future work.

\section{Expressiveness} \label{sec:eval}

To demonstrate the expressiveness of our grammar, we reproduce and extend two existing progressive visualization systems, ProReveal~\cite{Jo-ProReveal} and DimXplorer~\cite{DimXplorer-Desinging_progressive-and-interacitve-process-for-high-dimensional-data-analysis-2017-turkay}, and present a gallery of eleven working examples built with \ours.
We then discuss the coverage of \ours, characterizing the operations that fall outside the grammar.

\subsection{Reproducing the ProReveal System}
We reproduce ProReveal~\cite{Jo-ProReveal} based on its source code publicly available (\cref{fig:teaser}A).
ProReveal is a PVA system that provides two types of progressive visualizations: gradient plots for univariate
distributions and heatmaps for bivariate ones, both of which encode the
uncertainty of the current estimates.
It also supports safeguards, called \textit{PVA-Guards}, which are the user's expectations on the findings they made from a visualization, left on the visualization itself.
Once left, a PVA-Guard is continuously verified as the visualization updates, both in the middle of the computation and after it converges, and its validity is reported as a validity measure such as a $p$-value.

For reproduction, we replace the original client's Python WebSocket server
with \ours's API as the query backend.
The two progressive visualizations are expressed as binning specifications
(\cref{tab:proreveal_program_mapping}-1); for example, a bar chart of movie
counts per genre is expressed by grouping records by \code{genre} and
maintaining a count together with its confidence interval.
When the user drills down into a subset of interest, such as movies in the Action genre, we translate the interaction into a filtering predicate (\cref{tab:proreveal_program_mapping}-2).
We map runtime operations such as pausing, resuming, terminating, and prioritizing a visualization to \ours interaction requests (\cref{tab:proreveal_program_mapping}-3 and \cref{tab:proreveal_program_mapping}-4). For example, ProReveal computes the query at the top of its ongoing list first and lets the user reorder the list, which we reproduce by assigning a higher priority to the corresponding query (\cref{tab:proreveal_program_mapping}-4).
The Alternate option, which interleaves computation across visualizations, corresponds to \ours's default chunked round-robin scheduling.

Developers can also build client-side functionality on top of the progressive results the engine returns. For example, we implement PVA-Guards as a client-side analysis that computes a $p$-value from the aggregate values and their standard errors, assuming a normal distribution (\cref{tab:proreveal_program_mapping}-5).

\subsection{Extending the DimXplorer System}

We reproduce and extend DimXplorer~\cite{DimXplorer-Desinging_progressive-and-interacitve-process-for-high-dimensional-data-analysis-2017-turkay}, which supports progressive analysis of high-dimensional data through an online $k$-means clustering view (\cref{fig:teaser}B-1) and a PCA view (\cref{fig:teaser}B-2).
Users define a subset of interest by selecting a cluster or brushing the PCA view, and compare it against the rest of the data across a map view (\cref{fig:teaser}B-3), a histogram or scatterplot (\cref{fig:teaser}B-4), and a difference view that ranks attributes by Cohen's $d$ (\cref{fig:teaser}B-5).
We express these views and interactions as \ours specifications (\cref{tab:dimxplorer_program_mapping}); specifically, brushing the PCA view or selecting a cluster becomes a filter or a reducer over the corresponding binning state (\cref{tab:dimxplorer_program_mapping}-5), and the cluster-colored map becomes a binning over spatial coordinates and cluster labels (\cref{tab:dimxplorer_program_mapping}-3).

We further extend the original system in two ways.
First, whereas the original DimXplorer provides a fixed set of linked views, we let users create views dynamically from fields they select.
Such a query specifies PCA coordinates, cluster labels, and other selectable fields as binning keys (\cref{tab:dimxplorer_program_mapping}-4); PCA coordinates can be binned at a resolution suitable for brushing, while cluster labels aggregate values within a selected cluster.
Because the binning state is retained on the server, a newly created view draws on the data already processed and shows a result immediately rather than starting over.
Second, we place the online $k$-means view and the PCA view in the same lockstep group so that they share a single progression.
The original paper notes that progressive views advancing at different rates fluctuate independently, which makes their intermediate results hard to compare.
Lockstepping keeps the two queries aligned on the same execution chunks, so their results are always exposed at the same progress. 

\begin{table*}[t]
\centering

\footnotesize

\begin{minipage}[t]{0.235\textwidth}
\vspace{0pt}
\raggedright

\textbf{1. The original Vega-Lite spec}

\vspace{0.5mm}

\begin{tcolorbox}[
    enhanced,
    width=\linewidth,
    colback=black!1,
    colframe=black!15,
    boxrule=0.35pt,
    arc=1mm,
    left=1.2mm,
    right=1.2mm,
    top=0.8mm,
    bottom=0.8mm
]

{\scriptsize\ttfamily
\begin{tabular}{@{}l@{}}
\synkw{const} spec = \{ \\
\hspace*{2mm}\synkey{mark}: \synstr{"bar"}, \\
\hspace*{2mm}\synkey{data}: \{\synkey{name}: \synstr{"Crimes"}\}, \\
\hspace*{2mm}\synkey{encoding}: \{ \\
\hspace*{4mm}\synkey{x}: \{\synkey{field}: \synstr{"type"}, \\
\hspace*{9mm}\synkey{type}: \synstr{"nominal"}\}, \\
\hspace*{4mm}\synkey{y}: \{\schematag{schemaRed}{aggregate}: \synstr{"count"}, \\
\hspace*{9mm}\synkey{type}: \synstr{"quantitative"}\} \\
\hspace*{2mm}\} \\
\}; \\[0.5mm]
\synfun{vegaEmbed}(\synstr{"\#vis"}, spec);
\end{tabular}
}

\end{tcolorbox}

\end{minipage}
\hfill
%
\begin{minipage}[t]{0.235\textwidth}
\vspace{0pt}
\raggedright

\textbf{2. Derived \ours query}

\vspace{0.5mm}

\begin{tcolorbox}[
    enhanced,
    width=\linewidth,
    colback=black!1,
    colframe=black!15,
    boxrule=0.35pt,
    arc=1mm,
    left=1.2mm,
    right=1.2mm,
    top=0.8mm,
    bottom=0.8mm
]

{\scriptsize\ttfamily
\begin{tabular}{@{}l@{}}
\schematag{queryColor}{ProGramQuery}
\{ \\
\hspace*{3mm}\synkey{from}: \synstr{"Crimes"}, \\
\hspace*{3mm}\synkey{chunkType}: \synstr{"batch"}, \\
\hspace*{3mm}\synkey{ops}: [ \\
\hspace*{6mm}\{\synkey{type}:
    \schematag{binningColor}{"bin"}, \\
\hspace*{8mm}\synkey{id}: \synstr{"x"}, \\
\hspace*{8mm}\synkey{keys}: [ \\
\hspace*{11mm}\{\synkey{field}: \synstr{"type"}\}], \\
\hspace*{8mm}\synkey{values}: [ \\
\hspace*{11mm}\{\synkey{name}: \synstr{"count"}, \\ 
\hspace*{12mm}\synkey{op}: \synstr{"count"}\}] \\
\hspace*{6mm}\}] \\
\}
\end{tabular}
}

\end{tcolorbox}

\end{minipage}
\hfill
%
\begin{minipage}[t]{0.235\textwidth}
\vspace{0pt}
\raggedright

\textbf{3. Partial result}

\vspace{0.5mm}

\begin{tcolorbox}[
    enhanced,
    width=\linewidth,
    colback=black!1,
    colframe=black!15,
    boxrule=0.35pt,
    arc=1mm,
    left=1.2mm,
    right=1.2mm,
    top=0.8mm,
    bottom=0.8mm
]

{\scriptsize\ttfamily
\begin{tabular}{@{}l@{}}
\schematag{schemaRed}{System Response}\{ \\
\hspace*{1mm}\synkey{progress}: \{ \\
\hspace*{4mm}\synkey{processedRows}: \synnum{60000}, \\
\hspace*{4mm}\synkey{totalRows}: \synnum{7846809} \}, \\
\hspace*{1mm}\synkey{result}: [ \\
\hspace*{2mm}\{\synkey{key}: \{\synstr{"type"}: \synstr{"THEFT"}\}, \\
\hspace*{4mm}\synkey{values}: \{\synkey{count}: \synnum{12466}\}\}, \\
\hspace*{2mm}\{\synkey{key}: \{\synstr{"type"}: \synstr{"BATTERY"}\}, \\
\hspace*{4mm}\synkey{values}: \{\synkey{count}: \synnum{9263}\}\}, \\
\hspace*{2mm}\{\synkey{key}: \{\synstr{"type"}: \synstr{"ASSAULT"}\}, \\
\hspace*{4mm}\synkey{values}: \{\synkey{count}: \synnum{6543}\}\} \\
\hspace*{1mm}] \\
\}
\end{tabular}
}

\end{tcolorbox}

\end{minipage}
\hfill
%
\begin{minipage}[t]{0.235\textwidth}
\vspace{0pt}
\raggedright

\textbf{4. Visualizing the result}

\vspace{0.5mm}

\begin{tcolorbox}[
    enhanced,
    width=\linewidth,
    colback=black!1,
    colframe=black!15,
    boxrule=0.35pt,
    arc=1mm,
    left=1.2mm,
    right=1.2mm,
    top=0.8mm,
    bottom=0.8mm
]

{\scriptsize\ttfamily
\begin{tabular}{@{}l@{}}
pg.\synfun{on}(Q, (res) => \{ \\
\hspace*{1mm}\synfun{vegaEmbed}(\synstr{"\#vis"}, \{ \\
\hspace*{2mm}...spec, \\
\hspace*{2mm}\synkey{data}: \{ \\
\hspace*{3mm}\synkey{values}: res.\synfun{toRecords}()\}, \\
\hspace*{2mm}\synkey{encoding}: \{ \\
\hspace*{3mm}...spec.encoding, \\
\hspace*{3mm}\synkey{y}: \{\schematag{schemaRed}{field}: \synstr{"count"}, \\
\hspace*{6mm}\synkey{type}: \synstr{"quantitative"}\} \\
\hspace*{2mm}\} \\
\hspace*{1mm}\}); \\
\});
\end{tabular}
}

\end{tcolorbox}

\end{minipage}

\caption{Serving a Vega-Lite specification progressively with \ours.
The translator reads the original specification (1) and derives a \ours query that maps the categorical field to a binning key and the aggregate to a value definition (2).
The engine streams partial results (3), which are bound to the \code{data} of the original specification so that each update re-renders the same chart with more data (4).
}
\label{tab:vegalite_program_conversion}

\end{table*}

\subsection{The \ours Gallery}

We also showcase the \ours gallery consisting of eleven working examples and their specifications.
We highlight four notable examples among them: (1) visualizing the $t$-SNE result, (2) lockstepping between spatial heatmap and bar chart, (3) the forking example using two different bar charts, and (4) a rule-based translator to demonstrate how \ours can work with an existing visualization grammar, Vega-Lite~\cite{vega-lite}. 
All examples are available in the supplementary material.

First, we demonstrate iteration-based chunking with a query that runs $t$-SNE as a transforming operation over chunks of gradient-descent iterations (\cref{fig:teaser}C-1).
Before progressive execution begins, the engine loads all records to be embedded and runs a preparation step using openTSNE~\cite{openTSNE-2024-policar}: it constructs an approximate $k$-nearest-neighbor graph, computes pairwise affinities for the target perplexity, and initializes the embedding.
Each execution chunk then runs \code{iterSize} gradient-descent iterations and returns the current two-dimensional embedding, which the client uses to update the scatterplot.
The preparation step itself is not chunked, since this example aims to demonstrate iteration-based chunking rather than to integrate a particular progressive $t$-SNE implementation.
We leave such integration as future work.

Second, we demonstrate lockstepping between two visualizations, a spatial heatmap and a bar chart (\cref{fig:teaser}C-2).
The two visualizations are specified as separate queries over the same data source, placed in the same lockstep group.
The server therefore loads each execution chunk once and allows both queries to process the shared chunk.
The example also lets the user detach the two queries from the lockstep group and rejoin them during execution.
Detached queries continue independently from their current progress, and on rejoining, the engine realigns them before resuming shared execution.

Third, we demonstrate forking with two progressive bar charts.
The first chart shows the average fare for each pickup area, and its query maintains the total fare and the number of trips per area as its binning state.
While this chart is still progressing, the user creates a second chart that compares pickup areas by the number of trips.
Because the new query needs the same pickup-area bins and can derive its result from the trip counts already maintained, the engine automatically forks that binning state instead of starting over.
The second query therefore begins at the progress the first has reached, and the two then advance independently.

Finally, we demonstrate how the client-side specification of \ours lets developers reuse existing visualization specifications for progressive visualization (\cref{tab:vegalite_program_conversion}).
We provide a heuristic, rule-based translator that derives a \ours query from a Vega-Lite specification.
For example, a bar mark whose \code{x} is a nominal or ordinal field and whose \code{y} is an aggregate is read as requiring a binning operation over the \code{x} field, from which the translator maps the \code{x} field to a binning key and the \code{y} aggregate to a value definition (\cref{tab:vegalite_program_conversion}-2).
The translator does not rewrite the Vega-Lite specification; it produces the \ours query that serves it.
The original specification is reused as it is, with its \code{data} bound to the partial results streamed from the engine and its \code{y} encoding pointing at the corresponding value in those results, so each update re-renders the same chart with more data (\cref{tab:vegalite_program_conversion}-4).

\subsection{Grammar Coverage and Limitations}
We took the systems in our survey (\cref{table:pva_dev_survey}) as the reference set for what \ours should cover.
Every analytical operation that appears in more than two surveyed systems is available as a primitive, such as selection and filtering, count and sum, minimum and maximum, clustering, and projection.
The same holds for progressive control, where \ours provides concurrent queries, pausing and resuming, scheduling, and query prioritization.

Four types of operations, however, are not currently supported.
The first is operations that prioritize certain data.
\ours prioritizes queries but not the records within them, so it cannot express processing part of the data first, such as the higher-priority nodes of a hierarchy, or searching for a specific pattern before the rest of the data as PPMT~\cite{PPMT-Progressive-sequential-pattern-mining-Raveneau} does.
The second is domain-specific computation.
Seeking reusability, we chose the primitives from the operations that recur across the surveyed systems; incorporating operations bound to a single domain, such as shortest-path search in ProBGP~\cite{Ulmer-ProBGP}, remains future work.
The third is progressive algorithms we have not yet implemented.
DBSCAN, for example, has incremental variants that update clustering as new data arrive, and adding one as a transformer requires extending the engine rather than the grammar.
The fourth arises from abstracting an operation as a primitive, which leaves its internals out of reach.
A developer who wants to inspect or steer the $k$NN graph inside a $t$-SNE transformer, for example, cannot do so through a specification, as the engine keeps it hidden.


\section{Conclusion}

We present \ours, a grammar for specifying progressive computation pipelines for progressive visual analytics.
Rather than requiring developers to implement progressive execution and control mechanisms separately for each application, \ours expresses them through reusable query specifications informed by our survey.
The grammar supports both batch- and iteration-based progression, composable analytical operations, persistent binning states, and runtime control over concurrent queries.
It further provides mechanisms such as lockstepping, forking, and temporary prioritization to coordinate and reuse computation across progressive queries.
We demonstrate the expressiveness of \ours by reproducing ProReveal and DimXplorer, extending DimXplorer with dynamically created views and coordinated progression, and providing a gallery of additional progressive computation pipelines.

\bibliographystyle{abbrv-doi-hyperref}

\bibliography{template}

\appendix 

\end{document}